\documentclass[useAMS,usenatbib]{mn2e}

\usepackage{amssymb}
\usepackage{graphicx}

\title[X-ray/radio observations of Circinus X-1]{Linking jet emission and X-ray properties in the peculiar neutron star X-ray binary
Circinus X-1}
\author[P. Soleri et al.]{Paolo Soleri$^{1}$\thanks{E-mail:
p.soleri@uva.nl}, Valeriu Tudose$^{1,2,3,4}$, Rob Fender$^{5,1}$, Michiel van der Klis$^{1}$ and \newauthor Peter G. Jonker$^{6,7}$\\ 
$^{1}$Astronomical Institute `Anton Pannekoek', University of
Amsterdam, Science Park 904, 1098 XH, Amsterdam, the Netherlands\\
$^{2}$Astronomical Institute of the Romanian Academy, Cutitul de Argint 5, RO-040557 Bucharest, Romania\\
$^{3}$Research Center for Atomic Physics and Astrophysics, Atomistilor 405, RO-077125 Bucharest, Romania\\
$^{4}$Netherlands Institute for Radio Astronomy, Oude Hoogeveensedijk 4, 7991 PD Dwingeloo, the Netherlands\\
$^{5}$School of Physics and Astronomy, University of Southampton, Hampshire, SO17 1BJ, UK\\
$^{6}$SRON, Netherlands Institute for Space Research, 3584 CA Utrecht, the Netherlands\\
$^{7}$Harvard-Smithsonian Center for Astrophysics, 60 Garden Street, Cambridge, MA 02138, USA\\
}
\begin{document}

\date{Accepted 2009 June 23.  Received 2009 June 22; in original form 2009 May 4}

\pagerange{\pageref{firstpage}--\pageref{lastpage}} \pubyear{2009}

\maketitle

\label{firstpage}

\begin{abstract}
We present the results of simultaneous X-ray and radio observations of the peculiar Z-type neutron star X-ray binary Cir X-1, observed with
the Rossi X-ray timing explorer satellite and the Australia Telescope Compact Array in 2000 October and 2002 December. We
identify typical Z source behaviour in the power density spectra as well as characteristic Z patterns drawn in an X-ray hardness-intensity
diagram. Power spectra typical of bright atoll sources have also been identified at orbital phases after the periastron passage, while orbital
phases before the periastron passage are characterized by power spectra that are typical neither of Z nor of atoll sources.
We investigate the coupling between the X-ray and the radio properties, focusing on three orbital phases when
an enhancement of the radio flux density has been detected, to test the link between the inflow (X-ray) and the outflow (radio jet) to/from
the compact object. In two out of three cases we associate the presence of the radio jet to a spectral transition in the X-rays, although
the transition does not precede the radio flare, as detected in other Z sources. An analogous behaviour has recently been found in the black hole
candidate GX 339-4. In the third case, the radio light curve shows a similar shape to the X-ray light curve. We discuss
our results in the context of jet models, considering also black hole candidates.
\end{abstract}


\section{Introduction}
Low magnetic-field neutron star X-ray binaries (NSXBs) are often divided in two classes,
according to their correlated spectral and timing properties: the ``atoll'' and the ``Z''
sources (Hasinger \& van der Klis 1989, see van der Klis 2006 for a review), named after the shape of the track
they draw in X-ray color-color diagrams (CDs) and hardness-intensity diagrams (HIDs).

Z sources are brighter than atoll sources and are believed to accrete at
near-Eddington luminosities (0.5-1.0 $L_{Edd}$, van der Klis 2006). The mass accretion rate $\dot{m}$ is
thought to be the main parameter responsible for the differences between Z sources and atoll sources but the NS magnetic
field strength might also play a determinant role (Hasinger \& van der Klis 1989). To date seven Galactic
NSXBs have been classified as Z sources: six of them are persistent,
one is transient (XTE J1701-462, Homan et al. 2007a; the source evolved to atoll-like behaviour
at the end of its outburst; Homan et al. 2007b, Lin et al. 2009). They are
characterized by three-branched tracks in their CDs and HIDs that in some cases resemble the
character ``Z'', in other cases they have a more ``$\nu$-like'' shape. The branches are
called horizontal branch (HB), normal branch (NB) and flaring branch (FB) and identify three different states.
The mass accretion rate $\dot{m}$ is assumed to be the major physical parameter driving the transitions
between the states, increasing monotonically from the HB to the FB.
Motion along the branches of the ``Z'' usually takes place on time scales of hours to days while on
longer time scales it has been observed that the entire Z track can change its location in the CD and
HID or even its morphology (Cyg X-2, Kuulkers et al. 1996 and Wijnands et al. 1997; XTE J1701-422,
Homan et al. 2007a). Another possibility is that the mass accretion rate $\dot{m}$ drives the long-time scale
evolution between different systems (Z source Z-like shape, Z source $\nu$-like shape, atoll source)
while state transitions are mainly driven by disc instabilities (XTE J1701-462, Lin et al. 2009).

In timing studies, Z sources show a rich phenomenology of quasi periodic
oscillations (QPOs) and noise components, whose presence and properties are strongly correlated with
the position of the source along the Z track (van der Klis 2006). Three types of low-frequency QPOs
($<$100 Hz) are seen in the Z sources; they derive their name from the branches on which they were
originally found: horizontal branch (HBOs), normal branch (NBOs) and flaring branch oscillations (FBOs).
Twin kHz QPOs have also been detected in all the Z sources (marginally in XTE J1701-462,
Homan et al. 2007a).

All the Z-type NS sources are detected in radio (at variance with atoll sources, see Migliari \& Fender 2006),
showing large amplitude variability and both optically thin and optically thick emission. Penninx et al. (1988),
for the first time, found (in GX 17+2) that the radio emission varies as
a function of the position in the X-ray CD, decreasing with increasing mass accretion rate from the HB
(strongest radio emission) to the FB (weakest radio emission).
Hjellming et al (1990a, 1990b) found a consistent behaviour both in Cyg X-2 and in
Sco X-1, that according to Migliari \& Fender (2006) could be universal (but see Tan et al. 1992 for GX~5-1).
Migliari et al. (2007) observed GX 17+2 simultaneously in X-ray and radio and confirmed previous results, reporting a delay of
approximately $\sim 2$ hours between X-ray spectral transitions and the radio jet activity.
A similar behaviour has been found in the bright atoll source GX 13+1 (Homan et al. 2004), in which a change in the X-ray spectral
hardness precedes the beginning of the radio flares by approximately 40 minutes. The transient Z source  XTE J1701-462 has also been
detected in radio (Fender et al. 2007) when on the HB and NB.
Extended radio jets have been spatially resolved for Sco X-1 (Fomalont et al. 2001a) and have also
been associated with ultra-relativistic ejections (Fomalont et al. 2001b).
\begin{figure*}
\begin{tabular}{c}
\resizebox{14cm}{!}{\includegraphics{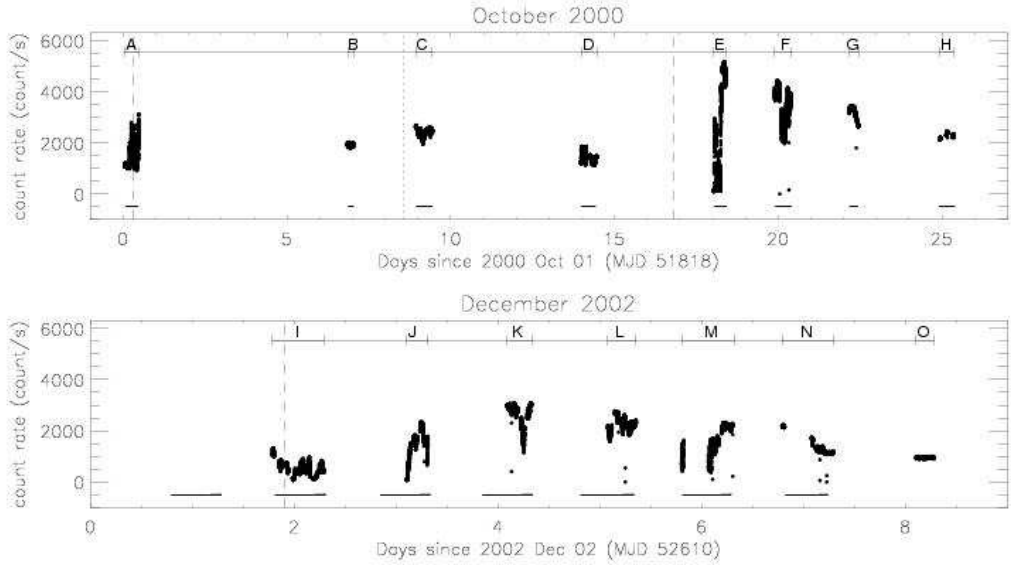}}
\end{tabular}
\caption{2-18 keV RXTE/PCA light curves for the two data sets examined here. Bin size is 16
seconds. Only counts from PCU2 were used. Horizontal lines in the bottom part of each light curve correspond to ATCA radio
observations. The letters A-O indicate 15 intervals in which we divided the data (see \S\ref{par:licu_hid}).
See Figures \ref{fig:HID_Oct} and \ref{fig:HID_Dec} for the corresponding hardness-intensity diagrams.
Vertical dashed lines mark the periastron passage (phase 0.0 of the orbital cycle), vertical dotted lines mark the apastron passage (phase 0.5
of the orbital cycle).}
\label{fig:licu_PCA}
\end{figure*}

\subsection{Cir X-1}
Cir X-1 is a peculiar X-ray binary discovered by Margon et al. (1971) characterized by flares
with a period of 16.55 days, observed first in the X-ray band (Kaluzienski et al. 1976) and then in the
infrared (Glass 1978; Glass 1994), radio (Haynes et al. 1978) and optical bands (Moneti 1992): these flares have
been interpreted as due to enhanced accretion close to the periastron passage of a highly eccentric
binary orbit ($e \, \sim \, 0.8$, Murdin et al. 1980, Nicolson, Glass \& Feast 1980).
The evidence that the system harbours a neutron star comes from the detection of type-I X-ray bursts
(Tennant et al. 1986a,b) and the more recent detection of twin kHz QPOs in the X-ray power density spectra
(Boutloukos et al. 2006). However Cir X-1 is a very peculiar neutron star system, since it features
several black-hole like properties. It is the most radio loud neutron star system (Whelan et al. 1977,
Haynes et al. 1978; Cyg X-3 and SS433 are brighter in radio but their nature is unclear), showing extended
jets on arcmin scale (Stewart et al. 1993; Tudose et al. 2006) that probably inflated the radio nebula in which the
system is embedded. It also shows arcsec-scale ultra-relativistic radio jets (Fender et al. 1998, 2004;
Tudose et al. 2008). On arcmin scale, X-ray jets have been recently detected
(Heinz et al. 2007, Soleri et al. 2008b): Cir X-1 is the only secure neutron star system in which
such ejections have been spatially resolved. 
The source is also characterized by a hard X-ray emission (Iaria et al. 2001, Ding et
al. 2003) and very strong X-ray variability (Jones et al. 1974, Samimi et al. 1979, Oosterbroek et al. 1995),
showing dramatic evolution of its X-ray luminosity, spectra and timing properties on timescales from
milliseconds to decades (Shirey et al. 1999, Parkinson et al. 2003).
Oosterbroek et al. (1995) identified atoll-source behaviour (strong band-limited noise
and non Z-like pattern in the CDs and HIDs) in the low-flux phases of Cir X-1 while Shirey et
al. (1998), in an analysis of Rossi X-ray Timing Explorer (RXTE) data, identified the first
typical Z-source features in the power density spectra
and then (Shirey et al. 1999) a complete Z track in its high-luminosity orbital phases.
Based on the properties of the type-I X-ray bursts, Jonker \& Nelemans (2004) estimated a distance to
the source of 7.8-10.5 kpc; according to a different measure of the galactic absorption, Iaria et al.
(2005) derived a smaller distance of 4.1 kpc (but see Jonker et al. 2007). The identification of the companion
star is still a matter of debate: from optical and infrared observations Johnston et al. (1999)
suggested a sub-giant companion star of 3-5 $M_{\odot}$, while Jonker et al. (2007), from analysis of the
optical spectrum, found evidence for super-giant characteristics.

\section{Observations and data analysis}
We analysed simultaneous X-ray (RXTE) and radio observations from the Australian Telescope Compact Array (ATCA)
performed during 2000 October and 2002 December. A log of the complete data set can be found in
Table \ref{tab:log_obs}. Orbital phases (both for RXTE and ATCA observations) are estimated using the radio ephemeris from Nicolson (2007).
These ephemeris mark the onset of the radio flare, not the peak. In Table \ref{tab:log_obs}, the first digit of the orbital phases increases
after every periastron passage.
\begin{table*}
\centering
\caption{Log of the RXTE and ATCA observations analysed in this paper. The data are divided in intervals (see also Figure
\ref{fig:licu_PCA} for the X-ray PCA light curve) according to criteria explained in \S\ref{par:licu_hid}. The first digit of the orbital phases
increases after every periastron passage. For more details on the ATCA observations we refer the reader to Table 1 in Tudose et al. (2008).}
\label{tab:log_obs}
\begin{tabular}{c c c c c c c}
\hline
\hline
\multicolumn{7}{c}{\bf{2000 October}}\\
\multicolumn{1}{c}{} & \multicolumn{1}{c}{Day of} & \multicolumn{3}{c}{} & \multicolumn{2}{c}{Orbital phases}\\
Interval & the month & RXTE beginning (MJD)$^{\mathrm{a}}$ & RXTE exposure$^{\mathrm{b}}$ (s) & ATCA exposure (h) & X-ray  & Radio \\
\hline
A        & 01     & 51818.034	         & 25875	     & 8.8		& 0.98-1.01    & 0.99-1.01\\
B        & 07-08  & 51824.851            & 5568	             & 4.0		& 1.40-1.41    & 1.40-1.41\\
C        & 09-10  & 51826.930            & 28155	     & 11.5		& 1.52-1.54    & 1.52-1.55\\
D        & 14-15  & 51831.977            & 22092             & 10.0	        & 1.83-1.86    & 1.83-1.85\\
E        & 19     & 51836.029            & 20475             & 8.8		& 2.07-2.10    & 2.08-2.10\\
F        & 20-21  & 51837.888	         & 23491	     & 11.5		& 2.19-2.22    & 2.19-2.22\\
G        & 23     & 51840.168	         & 15859	     & 6.4		& 2.32-2.34    & 2.32-2.34\\
H        & 25-26  & 51842.932	         & 9416	             & 11.5		& 2.49-2.52    & 2.49-2.52\\
\hline
\multicolumn{7}{c}{\bf{2002 December}}\\
\multicolumn{1}{c}{} & \multicolumn{1}{c}{Day of} & \multicolumn{3}{c}{} & \multicolumn{2}{c}{Orbital phases}\\
Interval & the month & RXTE beginning (MJD)$^{\mathrm{a}}$ & RXTE exposure$^{\mathrm{b}}$ (s) & ATCA exposure (h) & X-ray  & Radio \\
\hline
-        & 02    & -                     &   -               & 11.5              & -            & 0.93-0.96\\
I        & 03-04 & 52611.786             & 24002             & 11.9              & 0.99-1.02    & 0.99-1.02\\
J        & 05    & 52613.102             & 12423             & 11.5              & 1.07-1.09    & 1.06-1.09\\
K        & 06    & 52614.090             & 14728             & 11.9              & 1.13-1.15    & 1.12-1.15\\
L        & 07    & 52615.078             & 15520             & 12.3              & 1.19-1.21    & 1.18-1.21\\
M        & 07-08 & 52615.802             & 16336             & 11.5              & 1.24-1.27    & 1.24-1.27\\
N        & 08-09 & 52616.789             & 13920             & 10.0              & 1.30-1.33    & 1.30-1.32\\
O        & 10    & 52618.105             & 10735             & -                 & 1.38-1.39    & -        \\
\hline
\hline
\end{tabular}
\begin{list}{}{}
\item[$^{\mathrm{a}}$] Beginning of the observation in the ``Standard 2''-mode data
\item[$^{\mathrm{b}}$] Net exposure, excluding the gaps in between different spacecraft orbits and observations
\end{list}
\end{table*}

\subsection{X-ray data} \label{par:X_data}
We analysed 22 RXTE/PCA (proportional counter array, Jahoda et al. 1996) observations made between 2000
October 1 and 2000 October 26 and 23 RXTE/PCA observations performed between 2002 December 3 and
2002 December 10.

Background subtracted light curves with a time resolution of 16 seconds were obtained from
the ``Standard 2''-mode data, covering the energy range 2-18 keV. We used only data from proportional
counter array unit (PCU) number 2, the most reliable of
the five PCUs, always on during all our observations. We applied dead-time corrections.
We also defined two X-ray colors, a hard color (HC) and a broad color (BC), as the ratio of the count rates
in the following energy bands: (13-18)/(8.5-13) keV (HC) and (6.3-13)/(2-6.3) keV (BC). The energy bands have been
defined according to Shirey et al. (1998), to make comparisons easier.
We used the BC to produce HIDs, with the intensity defined as the count rate in the 2-18 keV energy band and
the HC to plot color curves.

PDS were extracted from high time resolution data from all the active PCUs
summed together, using standard Fast Fourier Transform (FFT) techniques (van der Klis 1989,
1995). The data were dead-time and background corrected. The PCA data mode used to produce the PDS were different
for the data sets of 2000 October and 2002 December (Single Bit and Event data for the former and only Single Bit
for the latter, see Jahoda et al. 1996 for an explanation), hence we could not extract power spectra in the same
energy band. We chose the energy range $\sim$2-33 keV (absolute channel 0-78) for 2000 October
and $\sim$2-10 keV (absolute channels 0-23) for 2002 December.
We extracted PDS with the same Nyquist frequency of 8192 Hz on segments with lengths of 128 s. 
We averaged the resulting PDS over certain selections of points in the HID (see \S\ref{par:PDS}),
we rebinned them logarithmically and we also subtracted the Poissonian noise, including the very large events
(VLE) contribution (Zhang 1995, Zhang et al. 1995). The PDS were then normalized to square
fractional {\it rms} (see Belloni \& Hasinger 1990). The fitting was carried out using a combination of
Lorentzians (Nowak 2000, Belloni et al. 2002), including a power-law component  when needed to take into account the very
low frequency noise (VLFN).
Following Belloni et al. (2002), we will quote the frequency $\nu_{max}$ at which the Lorentzian attains its maximum in
$\nu P(\nu)$ representation and the quality factor $Q$, where $\nu_{max}=\nu_0(1+1/4Q^2)^{1/2}$ and $Q=\nu_0/2\Delta$.
$\nu_0$ is the Lorentzian centroid frequency and $\Delta$ is its half-width-at-half-maximum.
Errors on fit parameters were determined using $\Delta\chi^2$=1.
\begin{figure*}
\begin{tabular}{c}
\resizebox{18cm}{!}{\includegraphics{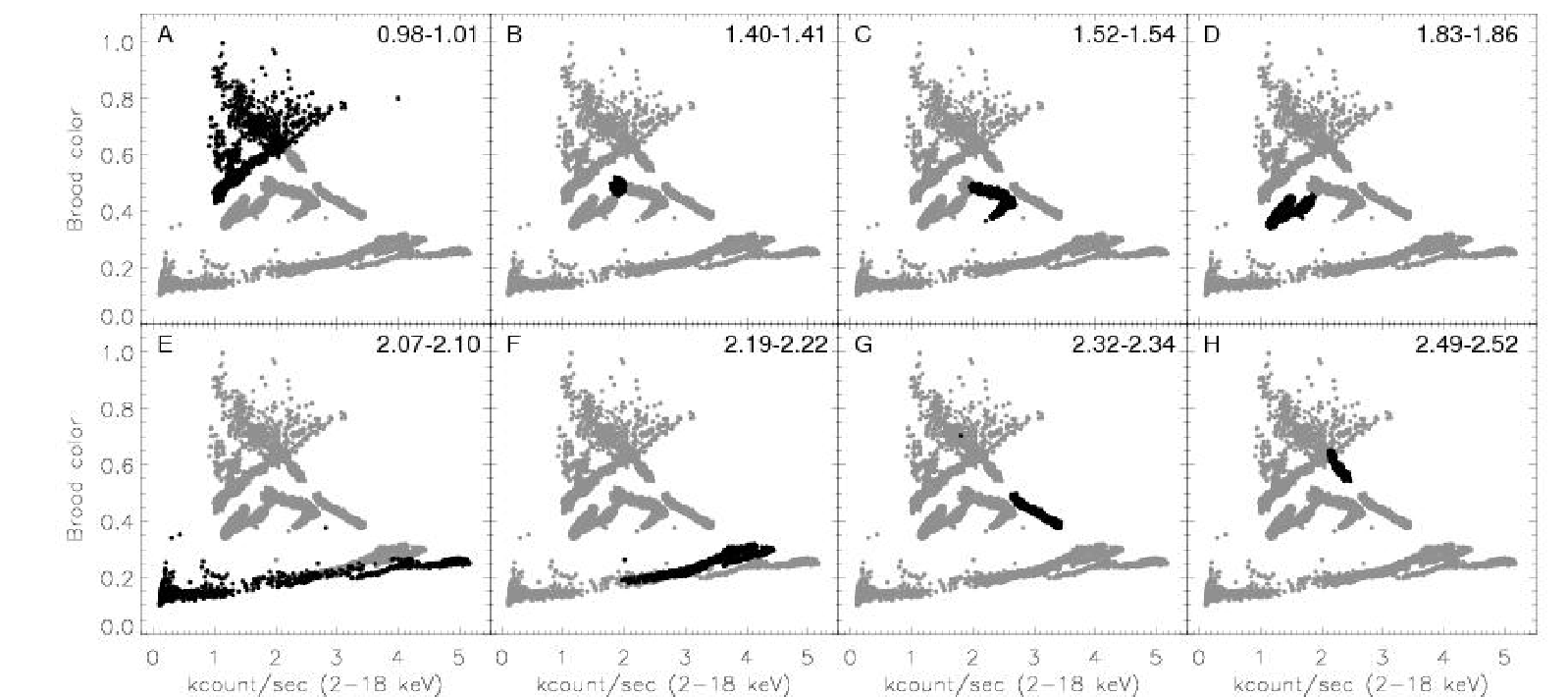}}
\end{tabular}
\caption{HID for the October 2000 data. Bin size is 16 seconds. Error bars are usually smaller then the symbols. In each panel the
points that correspond to one of the intervals listed in Table \ref{tab:log_obs} (from A to H only)
are marked in black, while the other points are plotted in grey. The name of the interval is written on the
top left part of the panels, while the corresponding orbital phase (from Table \ref{tab:log_obs}) is reported on the top-right part.}
\label{fig:HID_Oct}
\end{figure*} 
\begin{figure*}
\begin{tabular}{c}
\resizebox{18cm}{!}{\includegraphics{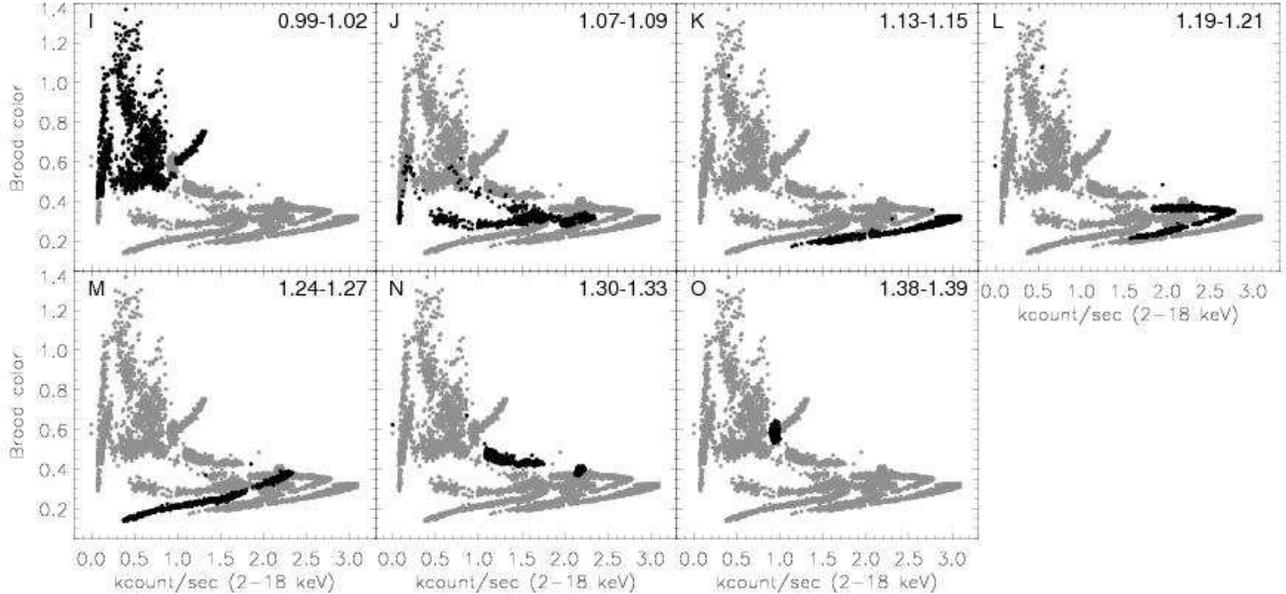}}
\end{tabular}
\caption{HID for the December 2002 data. Bin size is 16 seconds. Error bars are usually smaller then the symbols. In each panel the
points that correspond to one of the intervals listed in Table \ref{tab:log_obs} (from I to O only)
are marked in black, while the other points are plotted in grey. The name of the interval is written on the
top left part of the panels, while the corresponding orbital phase (from Table \ref{tab:log_obs}) is reported on the top-right part.
Note the different scales on the axis than Figure \ref{fig:HID_Oct}.}
\label{fig:HID_Dec}
\end{figure*} 

\subsection{Radio data}
We have observed Cir X-1 over multiple epochs on 2000 October 
and 2002 December at 4.8 and 8.6 GHz using the Australia Telescope Compact Array (ATCA). We used 
as primary calibrator PKS J1939-6342 (PKS B1934-638) and as secondary calibrator PMN J1524-5903 
(B1520-58). Standard calibration techniques were employed using the MIRIAD software (Sault, Teuben 
\& Wright 1995). For more details on the radio analysis we refer the reader to Tudose et al. (2008).

\section{Results and interpretation}
In this section we will proceed as follows: first we analyse the X-ray light curves, the HIDs and
the PDS, identifying typical Z source features, then we focus on the correlated X-ray/radio behaviour at
two particular orbital phases (0.0 and 0.5, periastron and apastron respectively), when radio jets are detected.

\subsection{Light curves and evolution in the hardness-intensity diagrams} \label{par:licu_hid}
Figure \ref{fig:licu_PCA} shows the 2-18 keV PCA light curves of Cir X-1 for our data set: in 2000 October the source crossed the periastron
twice during the time interval in which our observations have been performed while in our December-2002 campaign Cir X-1 crossed the
periastron only once (the first digit of the orbital phases in Table \ref{tab:log_obs} increases after every periastron passage).\\
The X-ray light curve of Cir X-1 is highly variable and in the last decades it has been
characterized by periods of very low flux and periods in which strong re-brightening occurred. 
Since a discussion of its morphology is beyond the scope of this paper (see Parkinson et al. 2003), we just note
that the PCA count rate reached higher peaks in 2000 October than in 2002 December, without further commenting on this aspect.\\
In some light curves, rapid changes in the X-ray count rate occur on short time scales, up to $\sim$3600 c/s (PCU2 only) in
$\sim$17 min (a factor $\sim$7 in count rate) on 2000 October 19. Dips are also present when close to phase 0.0 (2000
October 19; 2002 December 3, 4 and 5) as previously reported (e.g. Ding et al. 2006).\\
We created a HID both for 2000 October and 2002 December: the source systematically changes
its position in the diagram according to the orbital phase, drawing various patterns that, when not close to the periastron passage,
do not overlap in the HID. We have divided the data set in different intervals, that are chosen in such a way that each of them
corresponds to an ATCA observation (except for interval O, since no radio observations
are available). RXTE observations that belong to the same interval are usually characterized by consistent patterns in the HID.
This method led to 15 intervals, labeled A to O, marked in the X-ray light curves in Figure
\ref{fig:licu_PCA}. The associated orbital phases are given in Table \ref{tab:log_obs}. Figures \ref{fig:HID_Oct} and \ref{fig:HID_Dec}
show the location of the intervals in the HIDs for 2000 October (intervals A-H) and 2002 December (intervals I-O), respectively.
From now on we will refer to the interval names without necessarily reporting the date.\\
The 2000-October data set covers most of the orbital phases while the 2002-December one focuses on the periastron passage, in the phase
interval 0.99-1.39. Excluding intervals B, C, D and H (for which there is no overlapping in the two data sets), the HIDs for 2000 October and
2002 December have comparable shape (note the different scale on the axis, due to the decreasing secular trend of Cir X-1 count rate).
Position and patterns in the HIDs are strongly dependent on the orbital phase and that is
particularly evident around phase 0.0: before/across the periastron passage the source draws a cloud in the top-left part of
the HID (intervals A, I and part of interval J). Subsequently Cir X-1 moves in the
bottom part of the HID, drawing nearly horizontal strips with strong variations in count rate (intervals E, F, J, K, L and M).
Intervals B, C, D, F, G, H, N, O correspond to orbital phases not close to the periastron passage. They do not overlap with the cloud and the
horizontal strips and they are characterized by smaller variations in count rate.

\subsection{Identification of Z-source features} \label{par:PDS}
\begin{figure}
\begin{tabular}{c}
\resizebox{8.5cm}{!}{\includegraphics{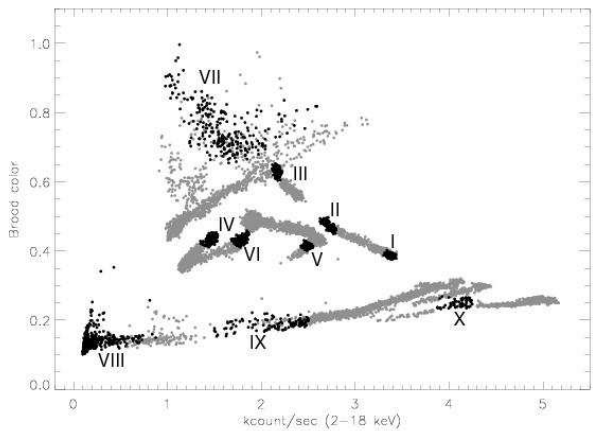}}
\end{tabular}
\caption{HID for the October 2000 data. Bin size is 16 seconds. Each Roman numeral refers to a region of points (marked in black) for which
we extracted and averaged a PDS, reported in Figure \ref{fig:pds_typical}. Grey dots represent points for which we do not show any PDS.}
\label{fig:hid_pds}
\end{figure}

\begin{figure*}
\begin{tabular}{c}
\resizebox{18cm}{!}{\includegraphics{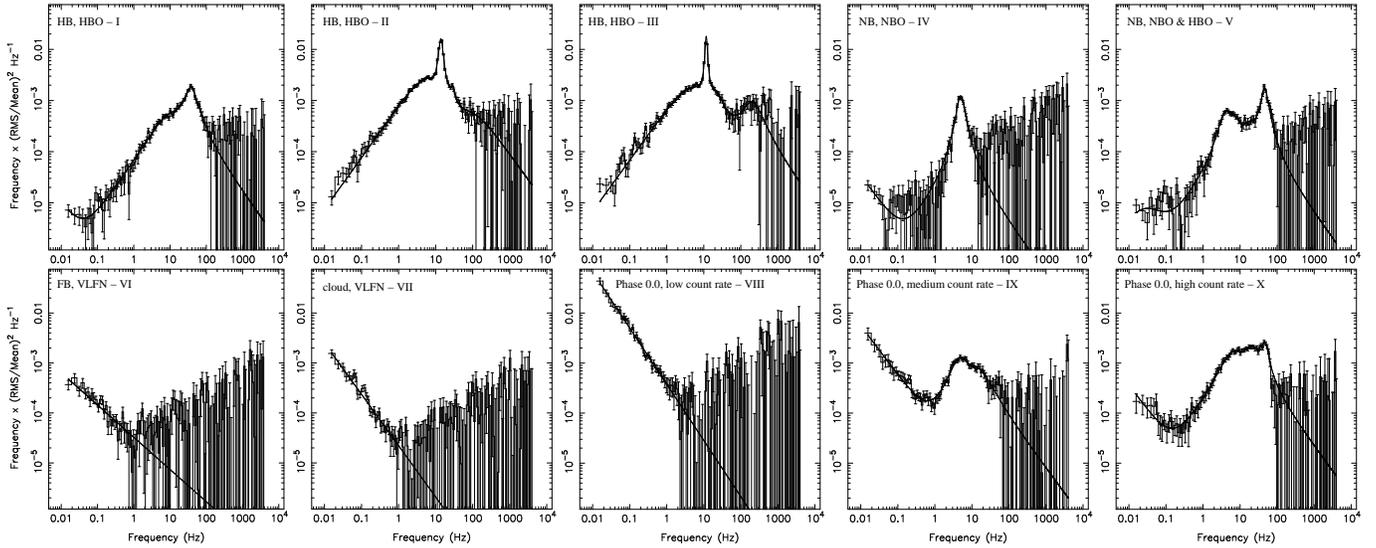}}
\end{tabular}
\caption{Average power density spectra calculated using data obtained in October 2000, selected according to the regions indicated in Figure
\ref{fig:hid_pds}. All power spectra are rms normalized and the Poissonian level has been subtracted. The solid line in each panel
shows the best fit to the spectrum. Best fit parameters are in Table \ref{tab:fit_pds}.}
\label{fig:pds_typical}
\end{figure*}

The aim of this section is to test if we can identify Z source features in the PDS and HIDs.
We focus on some observations performed in 2000 October, since the associated patterns in the HID
and PDS are representative of the source behaviour. We chose 2000 October since those data, differently from the ones from 2002 December,
cover all the orbital phases.\\
Figure \ref{fig:hid_pds} shows a HID for 2000 October. We marked 10 regions of points for which we have extracted and averaged
power spectra. Each region and its corresponding power spectrum (reported in Figure \ref{fig:pds_typical}) is marked
with a Roman numeral (I to X). We fitted the PDS as described in \S\ref{par:X_data} and we reported the best fit parameters 
in Table \ref{tab:fit_pds}. For the remaining part of this
section we will implicitly refer to Figures \ref{fig:hid_pds} and \ref{fig:pds_typical} and Table \ref{tab:fit_pds}.
\begin{table*}
\centering
\caption{Best fit parameters of the PDS in Figures \ref{fig:pds_typical} and \ref{fig:hid_licu_pds_Dec2002}.}
\label{tab:fit_pds}
\begin{tabular}{l c r r r r r}
\hline
Interval/PDS & Type & \multicolumn{1}{c}{$\nu_{max}$ (Hz)} &\multicolumn{1}{c}{$Q$} & rms (\%) & Significance ($\sigma$) & $\chi^{2}/dof$\\
\hline
\multicolumn{7}{c}{\bf{HB}}\\
G/PDS-I      & HBO & 37.67$\pm$0.0002 & 1.12$\pm$0.10 & 4.34$\pm$0.17 & 11.17                 &           130/114\\
G/PDS-II     & HBO & 13.86$\pm$0.03   & 3.54$\pm$0.09 & 7.95$\pm$0.08 & 46.52                 &           184/108\\
G/PDS-II     & HBO$^{\mathrm{a}}$ & 28.83$\pm$1.00 & 3.83$\pm$1.11 & 1.83$\pm$0.22 & 4.41     &  -$^{\mathrm{b}}$\\
H/PDS-III    & HBO & 11.65$\pm$0.02   & 6.98$\pm$0.31 & 6.06$\pm$0.07 & 43.34                 &           152/110\\
H/PDS-III    & HBO$^{\mathrm{a}}$ & 22.78$\pm$1.28 & 1.41$\pm$0.39 & 2.30$\pm$0.31 & 3.99     &  -$^{\mathrm{b}}$\\
H/PDS-III    & kHz & 201.16$\pm$26.42 & 0.63$\pm$0.22 & 3.82$\pm$0.32 & 6.26                  &  -$^{\mathrm{b}}$\\                   
\multicolumn{7}{c}{\bf{NB}}\\
D/PDS-IV     & NBO  & 4.99$\pm$0.05    & 1.88$\pm$0.09 & 3.04$\pm$0.05 & 33.42                 &           139/117\\
D/PDS-IV     & VLFN & 2.01$\pm$0.29$^{\mathrm{c}}$ & - & 0.59$\pm$0.04 & -                     & -$^{\mathrm{b}}$\\
C/PDS-V   &   NBO-first  & 7.20$\pm$0.56 & 1.32$\pm$0.48 & -$^{\mathrm{e}}$ & 2.70            &           150/108\\
C/PDS-V   & NBO-second & 4.05$\pm$0.17 & 1.18$\pm$0.12 & 2.93$\pm$0.0.41$^{\mathrm{f}}$ & 4.74 & -$^{\mathrm{b}}$\\
C/PDS-V      & HBO &  45.95$\pm$0.46   &  2.07$\pm$0.24 & 3.39$\pm$0.18           & 7.76      & -$^{\mathrm{b}}$ \\
\multicolumn{7}{c}{\bf{FB}}\\
D/PDS-VI     & VLFN & 1.64$\pm$0.05$^{\mathrm{c}}$ & - & 3.13$\pm$0.10$^{\mathrm{d}}$ & -      & 144/120         \\ 
\multicolumn{7}{c}{\bf{Cloud}}\\
A/PDS-VII     & VLFN & 2.03$\pm$0.04$^{\mathrm{c}}$ & - & 4.94$\pm$0.17$^{\mathrm{d}}$ & -      &  144/120         \\        
\multicolumn{7}{c}{\bf{Phase 0.0 - 2000 October - Horizontal strips}}\\
E/PDS-VIII  & VLFN & 2.12$\pm$0.03$^{\mathrm{c}}$ & - & 24.35$\pm$0.66$^{\mathrm{d}}$ & -      &  124/120        \\
E/PDS-IX    & VLFN & 1.98$\pm$0.05$^{\mathrm{c}}$ & - & 7.56$\pm$0.33$^{\mathrm{d}}$  & -      &  140/114        \\
E/PDS-IX    &  -   & 4.66$\pm$0.19   & 0.77$\pm$0.08 & 4.07$\pm$0.27                & 6.94     & -$^{\mathrm{b}}$\\         
E/PDS-IX    &  -   & 14.55$\pm$1.66  & 0.41$\pm$0.14 & 3.57$\pm$0.34                & 5.70     & -$^{\mathrm{b}}$\\
E/PDS-X     & VLFN & 2.00$\pm$0.15$^{\mathrm{c}}$ & - & 1.91$\pm$0.16$^{\mathrm{d}}$ & -       &  95.7/111       \\ 
E/PDS-X    &  -   & 6.56$\pm$0.43   & 0.47$\pm$0.05 & 5.34$\pm$0.28               & 9.01      & -$^{\mathrm{b}}$\\
E/PDS-X    &  -   & 21.17$\pm$1.07  & 0.75$\pm$0.15 & 5.00$\pm$0.43               & 6.15      & -$^{\mathrm{b}}$\\
E/PDS-X    &  -   & 46.76$\pm$0.88  & 2.28$\pm$0.33 & 3.54$\pm$0.27               & 6.70      & -$^{\mathrm{b}}$\\
\multicolumn{7}{c}{\bf{Phase 0.0 - 2002 December}}\\
I/PDS-XI & VLFN & 2.18$\pm$0.05$^{\mathrm{c}}$ & - & 6.69$\pm$0.26$^{\mathrm{d}}$   & -        &    120/120      \\
J/PDS-XII  & VLFN & 1.82$\pm$0.10$^{\mathrm{c}}$ & - & 3.74$\pm$0.20$^{\mathrm{d}}$ & -        &    110/111      \\
J/PDS-XII  &   -  & 4.18$\pm$0.14  & 1.56$\pm$0.35   & 2.57$\pm$0.33                &   4.60   & -$^{\mathrm{b}}$\\
J/PDS-XII  &   -  & 13.01$\pm$2.22   & 0.22$\pm$0.15 & 4.21$\pm$0.35                &  5.66    & -$^{\mathrm{b}}$\\
J/PDS-XII  &   -  & 51.63$\pm$1.57  & 2.29$\pm$0.60 & 3.15$\pm$0.31                 &  5.19    & -$^{\mathrm{b}}$\\
\hline 
\end{tabular}
\begin{list}{}{}
\item[$^{\mathrm{a}}$] Harmonic peak
\item[$^{\mathrm{b}}$] Already reported
\item[$^{\mathrm{c}}$] Index $\Gamma$ of the power law
\item[$^{\mathrm{d}}$] Integrated in the interval 0.1-100 Hz
\item[$^{\mathrm{e}}$] Peak fitted with two Lorentzians
\item[$^{\mathrm{f}}$] Sum of the rms of the two Lorentzians
\end{list}
\end{table*}
\subsubsection{Horizontal Branch}
Intervals G and H are characterized by a definite branch in the HID, on which the count rate decreases in time while the
value of the BC increases. PDS-I, -II and -III are extracted in three different positions on this branch. They show a highly
significant peak, the frequency of which correlates with the count rate, decreasing from $\nu \sim 38$ Hz to
$\nu \sim 12$ Hz while the quality factor $Q$ increases from 1.12$\pm$0.10 to 6.98$\pm$0.31. Its fractional rms
is in the range (4.34$\pm$0.17)\% - (7.95$\pm$0.08)\%. We identify the branch and the peak as a HB and a HBO, respectively. Their 
properties resemble those of the HB and the HBO identified in Shirey et al. (1998, 1999). PDS-I and PDS-II, in which the
HBO has a higher fractional amplitude, also show a harmonic peak. PDS-III features a significant broad kHz peak,
with $\nu = 201.16\pm26.42$ Hz and quality factor $Q = 0.63\pm0.22 $. No twin kHz QPOs have been detected in our data.

\subsubsection{Normal Branch} \label{par:normal_branch}
Interval D consists of two segments connected with a vertex. They both have an opposite inclination in the HID compared to intervals G and
H. We extracted power spectra in several positions along the branches and we noticed that, in one of them, they feature
(see PDS-IV) a peak with rather stable frequency around 5 Hz,
the centroid frequency of which does not correlate with the count rate. The peak is broader than the HBO (the one detected
in intervals G and H), with quality factor $Q \lesssim 2$ and with lower fractional rms ($\sim 3$\%). We identify the branch and the
peak as the NB and the NBO, respectively, in agreement with other identifications of NB and
NBO presented in Shirey et al. (1998, 1999). A power law is also needed to fit the VLFN in PDS-IV, with very low factional amplitude
(0.59$\pm$0.04)\%.\\ 
The segment of interval C in which we extracted PDS-V has the same inclination in the HID as the segment of interval D in which we extracted
PDS-IV. In PDS-V, the peak
at low frequency can be identified with a NBO, for the same reasons listed for PDS-IV. Fitting the peak with only one Lorentzian
gave a very poor fit: for this reason we used two Lorentzians, and in Table \ref{tab:fit_pds} we
report the parameters of both of them. The total fractional rms
has been calculated adding the rms of the two peaks. Although the significance of the first Lorentzian is $2.7 \sigma$, this component is
needed in order to get an acceptable fit. Another peak appears in PDS-V: its centroid frequency correlates with the count rate and its
quality factor $Q$ is $\sim$2: we identify this peak as a HBO. HBOs on the NB have already been detected in other
Z sources (GX 340+0, Penninx et al. 1991; GX 17+2, Wijnands et al. 1996, Homan et al. 2002; Cyg X-2, Wijnands et al. 1997; GX 5-1, Dotani
1988) but they have never been clearly detected in Cir X-1.

\subsubsection{Flaring Branch} \label{par:flaring_branch}
In \S\ref{par:normal_branch} we showed that one of the two branches of interval D can be identified as a NB. After the vertex the
inclination of the branch in the HID slightly changes and the source properties as well: in the light curve Cir X-1 shows bigger variations
in count rate than in the first segment and the PDS features a different kind of variability. PDS-VI has been extracted on this second part
of interval D. It only shows VLFN that can be fitted with a power law with spectral index $\Gamma=1.64\pm0.05$. Its fractional rms is
(3.13$\pm$0.10)\%. Considering both the pattern in the HID (this branch has a common vertex with a NB) and the PDS itself, dominated by VLFN,
we can identify this branch as a FB. A FB with the same properties has already been identified in Cir X-1
(Shirey et al. 1998, 1999) and in other Z sources (van der Klis et al. 2006) while a FBO has only been reported in Sco X-1 and GX 17+2
(Priedhorsky et al. 1986 and Homan et al. 2002, respectively) and possibly in GX 5-1 (Jonker et al. 2002).

\subsubsection{Cloud} \label{par:cloud}
Interval A was observed during 7 RXTE orbits in which the last 4 show a light curve characterized by rapid variations of the count rate
on a short time scale ($\sim$ 1300 c/s in PCU2 in $\sim$ 600 sec). PDS-VII was extracted on the fifth orbit and
it only shows VLFN that can be fitted with a power law with spectral index $\Gamma=2.03\pm0.04$. Its fractional rms is
4.94$\pm$0.17\%. The extraction region does not correspond to a definite branch in the HID but it rather resembles a random selection in a
cloud of points. Although the properties of the power spectrum might suggest that this region is a FB (but note the different slope of the
power law compared to the VLFN in \S\ref{par:flaring_branch}), the fact that no branch is drawn in the HID does not allow us to propose
such identification.

\subsubsection{Phase 0.0 - Horizontal strips} \label{par:phase0.0}
Interval E covers the orbital phases that follow the periastron passage. In the HID this interval draws a branch characterized by
a stable BC (in the range $\sim$0.1-0.25) and large changes in the count rate (0-5200 c/s, PCU2 only). Power density spectra
show strong correlation with the count rate, with characteristic frequencies moving towards high values with increasing count
rate.\\
PDS-VIII has been extracted in a region characterized by low count rate ($< 600$ c/s, PCU2) and frequent dips in which the PCA count rate
approaches $\sim0$ c/s. The power spectrum features strong VLFN that can be modeled using a power law with index $\Gamma = 2.12\pm0.03$ and
fractional rms (24.35$\pm$0.66)\%. PDS-IX is an example of "intermediate count rate" power spectrum. We extracted it in a region
where the count rate varies in the range 1500-2500 c/s (PCU2). The VLFN is still present but its fractional rms (7.56$\pm$0.33)\%
is $\sim 3$ times lower than in PDS-VIII. Two Lorentzians are needed in order to get a satisfactory fit: their 
$\nu$ are respectively 4.66$\pm$0.19 Hz and 14.55$\pm$1.66 Hz. The low-frequency one has the highest fractional amplitude.
PDS-X has been averaged in a high-count rate region in the HID ($\sim$ 4000 c/s). It shows a weak VLFN (fractional rms
1.91$\pm$0.16\%) and 3 Lorentzians are required to get an acceptable fit, with frequencies $\nu$ of 6.56$\pm$0.43 Hz,
21.17$\pm$1.07 Hz and 46.76$\pm$0.88 Hz respectively.\\
From the analysis of the power density spectra extracted on interval E it is evident that no HBO, NBO or VLFN are present.
Although the source draws rather regular branches in the HID, our timing studies show that Z branches can not be identified here: at the
orbital phases that follow the periastron passage the Cir X-1 does not behave as a Z source.
\begin{figure*}
\begin{tabular}{c}
\resizebox{17cm}{!}{\includegraphics{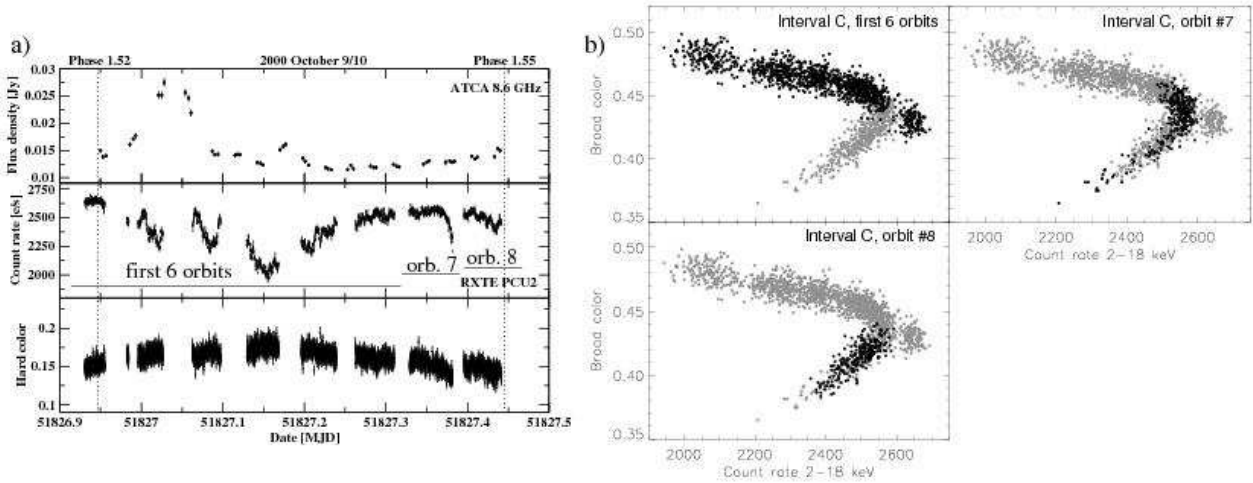}}
\end{tabular}
\caption{{\it a)} simultaneous ATCA (8.6 GHz, top panel) and RXTE (PCU2 only, middle panel) light curve of Cir X-1 obtained on 2000
October 9-10 (interval C). Horizontal lines mark different RXTE orbits. The HC curve is also plotted in the bottom panel. 
The bin size in the radio light curve is 5 min. Vertical dotted lines mark two orbital phases, as indicated. {\it b)} HIDs for interval C. 
In the three HIDs we marked in black the points corresponding to the first six, the seventh and the eighth RXTE orbits, respectively. The
other points are plotted in grey.}
\label{fig:licu_rad_hid_oct}
\end{figure*}

\subsection{Correlated X-ray/radio behaviour}
The detailed ATCA light curves and radio maps for 2000 October and 2002 December are reported in Tudose et al. (2008). The radio flux
density in 2002 December is systematically lower than in 2000 October. Since a complete analysis of these radio observations is already
described in Tudose et al. (2008), here we focus on the correlated X-ray/radio behaviour at two specific orbital phases: phase 0.5 in
2000 October (09-10) and phase 0.0 in 2000 October (01 and 19-21) and 2002 December (03-05).

\subsubsection{2000 October 09-10 - phase 0.5}  \label{par:0.5}
Figure \ref{fig:licu_rad_hid_oct} (a, top panel) shows the ATCA radio light curve at 8.6 GHz on 2000 October 09-10 (it corresponds to
interval C), in which a sequence of two radio flares with an interval of about 1 hour in between them is evident. These flares
occur after the apastron passage and they have been recently reported by Tudose et al. (2008).
While flares at phase 0.0 are a common feature that has been extensively studied, this is the first time that flares associated with other
orbital phases are clearly detected (Fender 1997 presented circumstantial evidence for flares at orbital phases different from 0.0).
The radio flux smoothly increases after the end of the second flare which might suggest that the flaring sequence did not end yet, although
the data do not allow to confirm or exclude this.\\
Figure \ref{fig:licu_rad_hid_oct} (a, middle and bottom panels) also shows the X-ray light and color curves for interval C. In the first
six RXTE orbits the source tracks a branch in in the HID (Figure \ref{fig:licu_rad_hid_oct}b)
that, by similarity with intervals G and H, we identify with a HB. The PDS extracted in this segment confirm the branch identification.
During the seventh RXTE orbit the source path in the HID makes a turn and in the RXTE-orbit 8 it is firmly
on another branch, that we already identified with the NB in \S\ref{par:normal_branch}. The identification of the branches is made robust
also by inspecting the HC curve: during the first six RXTE orbits the HC curve and the light curve anti-correlate while during
RXTE-orbits 6 and 7 they follow the same behaviour. The X-ray state transition occurs about 2.5-3 hours after the end of the second radio
flare.

\subsubsection{2002 December 03-05 - phase 0.0} \label{par:Dec2002}
In Figure \ref{fig:licu_rad_dec} (top panel) we reported the ATCA light curve at 8.6 GHz on 2002 December 03-06. On 2002 December 5
Cir X-1 shows a sequence of radio flares. We also show the RXTE light curve
and the HC curve (middle and bottom panel, respectively), in which one can see that the X-ray count rate increases at
the beginning of the flaring activity. From Figure \ref{fig:HID_Dec} it is evident that the source changes
its position in the HID from interval I to interval J: the former is characterized by a cloud of points with
count rate smaller than $\sim 1400$ c/s (PCU2) and BC above 0.4, the latter by horizontal strips with count rate up to
$\sim 2400$ c/s (PCU2) and BC below 0.4 (excluding few transition points from the cloud). Figure \ref{fig:hid_licu_pds_Dec2002}a
shows a zoom of the RXTE light curve and HID on intervals I and J. We averaged PDS-XI on the last orbit of interval I, in a region of
points located in the ``cloud'' (see Figure \ref{fig:hid_licu_pds_Dec2002}a, top-left and bottom-left panels). PDS-XII is averaged on the
first orbit of interval J, in a region that includes a few points from the
clouds and most of the points in the ``horizontal strips'' (see Figure \ref{fig:hid_licu_pds_Dec2002}a, top-right and bottom-right panels).
Both the extraction regions (in the light curve and in the HID) and the power
spectra are reported in Figure \ref{fig:hid_licu_pds_Dec2002}a,b. As we already noticed for region A and E in 2000 October (see \S
\ref{par:phase0.0}), the behaviour of
Cir X-1 in the light curve and in the HID completely changes from the cloud to the horizontal strips, and the properties of the power
spectrum change as well. Best fit parameters are reported in Table \ref{tab:fit_pds}. PDS-XI can be fitted only with a power law that takes
into account the VLFN. PDS-XII needs to be fitted with a power law and three Lorentzians, to fit the VLFN, two peaks at $\sim$4
and $\sim$52 Hz and broad noise in between them. The transition from the cloud to the horizontal strips area does not happen at
once: at the end of interval J, for a few minutes the count rate goes below 1000 c/s. That means that the source moves back to the
cloud, then Cir X-1 goes back to the horizontal strips.  

\subsubsection{2000 October 01 - phase 0.0}
Interval A and the simultaneous radio observation cover the periastron passage but they both end
at phase 1.01, when the radio flare is not observable yet (the next radio observation started at orbital
phase 1.40). By comparison, on 2002 December 03-05, the flare was observed only after phase $\sim 1.06$ (we stress again
that phase 0.0 marks the onset of the flare and not the peak). We will not further comment on observation A and its simultaneous radio
observation.

\subsubsection{2000 October 19-21 - phase 0.0}
\begin{figure}
\begin{tabular}{c}
\resizebox{9.5cm}{!}{\includegraphics[angle=-90]{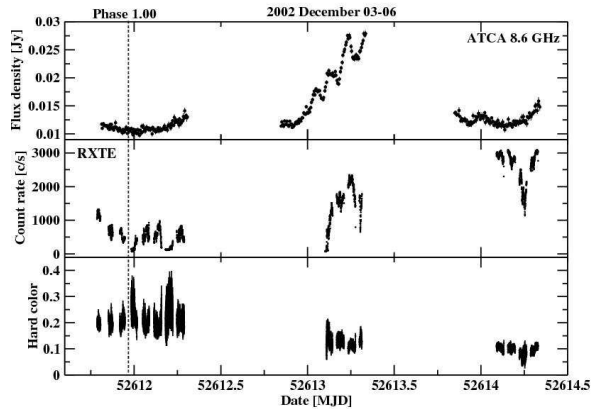}}
\end{tabular}
\caption{Simultaneous ATCA (8.6 GHz, top panel) and RXTE (PCU2 only, middle panel) light curve of Cir X-1 obtained on 2002
December 03-06 (intervals I, J and K). The HC curve is also reported in the bottom panel. The bin size in the radio light curve is 5 min.
The vertical dotted line marks phase 1.00 of the orbital cycle.}
\label{fig:licu_rad_dec}
\end{figure}
RXTE and ATCA observed Cir X-1 close to the passage through phase 0.0 of the orbital cycle also on 2000 October 19-21. The radio
light curve at 8.6 GHz is reported in Figure \ref{fig:radio_X_19_21_Oct} (top panel). When the radio observation started, the
flare was perhaps already on. The initial orbital phase is comparable to the starting phase
of the radio flare in 2002 December. The flaring activity is still detected during October 20-21, so it extends in time longer than in
2002 December. Some re-brightening occurred after the end of the main flare, with little bumps like in
2002 December. Figure \ref{fig:radio_X_19_21_Oct} also shows the X-ray light and color curve (middle and bottom panel, respectively)
for intervals E and F. Their position in the HID is shown in 
Figure \ref{fig:HID_Oct}, in which one can see that both intervals draw similar tracks in the diagram. The
only difference between them is the different shape of the light curve, rapidly variable in interval E and more stable
during interval F. No transition is observed in observations E and F. By comparison with intervals I and J (2002 December 03-05) it might be
that the transition already happened when observation E started. As we already showed in \S\ref{par:PDS}, the PDS behaviour during
these intervals is governed by the count rate rather than the source being on one of the Z branches. 
From Figure \ref{fig:radio_X_19_21_Oct} one can see that the radio and the X-ray light curve have a rather similar shape, and this
holds both for interval E and F. The color curve is characterized by oscillations during the first four orbits of interval E, when the count
rate is highly variable (between $\sim 0 - 3000$ c/s). Afterwards it stabilizes, correlating with the count rate for the remaining two
orbits of interval E and for the whole interval F.

\section{Discussion}
Cir X-1 features Z-source branches in the HIDs and characteristic Z-source noise components and QPOs in the power density spectra.
Some regions in the HID can not be identified in terms of Z branches.\\
HB and NB have been identified as well as the corresponding QPOs (PDS-I, -II, -III and -IV). The HBO has been also detected when the
source is on the the NB, together with a NBO (PDS-V). This has already been reported for other Z sources, GX 340+0 (Penninx et al. 1991),
GX 17+2 (Wijnands et al. 1996, Homan et al. 2002), Cyg X-2 (Wijnands et al. 1997) and GX 5-1 (Dotani 1988). We also identified the FB,
without any FBO but with VLFN (PDS-VI).\\
PDS-VII also shows VLFN but the extraction region (interval A) does not form a branch in the HID, so it is not a FB.
Intervals A and I (they display similar properties), that we named ``cloud'', can not be classified in terms of Z branches.\\ 
After the periastron passage the identification of Z patterns is difficult. The source draws branches in the HID (we named them
``horizontal strips'', intervals E, F, J, K, L and M) and in the power density spectra, noise components and QPOs shift towards higher
frequencies following a positive correlation with the count rate (like in the HB). Their properties are not reminiscent of any Z
branch. PDS-VIII is dominated by a strong VLFN and no QPO is present.
PDS-IX shows VLFN (although much weaker than in PDS-VIII) but two Lorentzians are also needed to fit it. PDS-X still shows VLFN (much
weaker than PDS-IX) and three Lorentzians are needed. Noise components extend to higher frequencies than PDS-IX ($\sim$47 Hz compared to
$\sim$15 Hz). The shape of PDS-IX and PDS-X is reminiscent of typical power density spectra of bright atoll sources that spend
most of their time in the soft state (the so called ``banana''): GX 9+9, GX 3+1, GX 9+1 and GX 13+1 (see van der Klis 2006 and
references therein). Their power spectra feature a broad high frequency noise (HFN, high compared to the VLFN) that usually needs to be
modeled with one or more Lorentzians (see Figures 1 and 8 in Schnerr et al. 2003). This is rather interesting: although the patterns that Cir
X-1 draws in the HID after the periastron passage are not atoll-like, power density spectra extracted there resemble atoll-source typical
power spectra in the lower-banana state. Note that the atoll behaviour becomes evident only after the periastron passage, in
a part of the orbital cycle characterized by a high accretion rate. This suggests that the mass accretion rate $\dot{m}$ is not the only
responsible (see Lin et al. 2009) for the changes between different Z and atoll sources.\\
Timing features typical of Z sources have already been identified along the whole orbit in Shirey et al. (1998). A complete
identification of the Z branches has been reported for the first time in Shirey et al. (1999), in which the observing campaign focused
around phase 0.0 of the orbital cycle. At variance with Shirey et al. (1998, 1999), we could not identify Z-source properties when close to
the periastron passage.
\begin{figure*}
\begin{tabular}{c}
\resizebox{17cm}{!}{\includegraphics{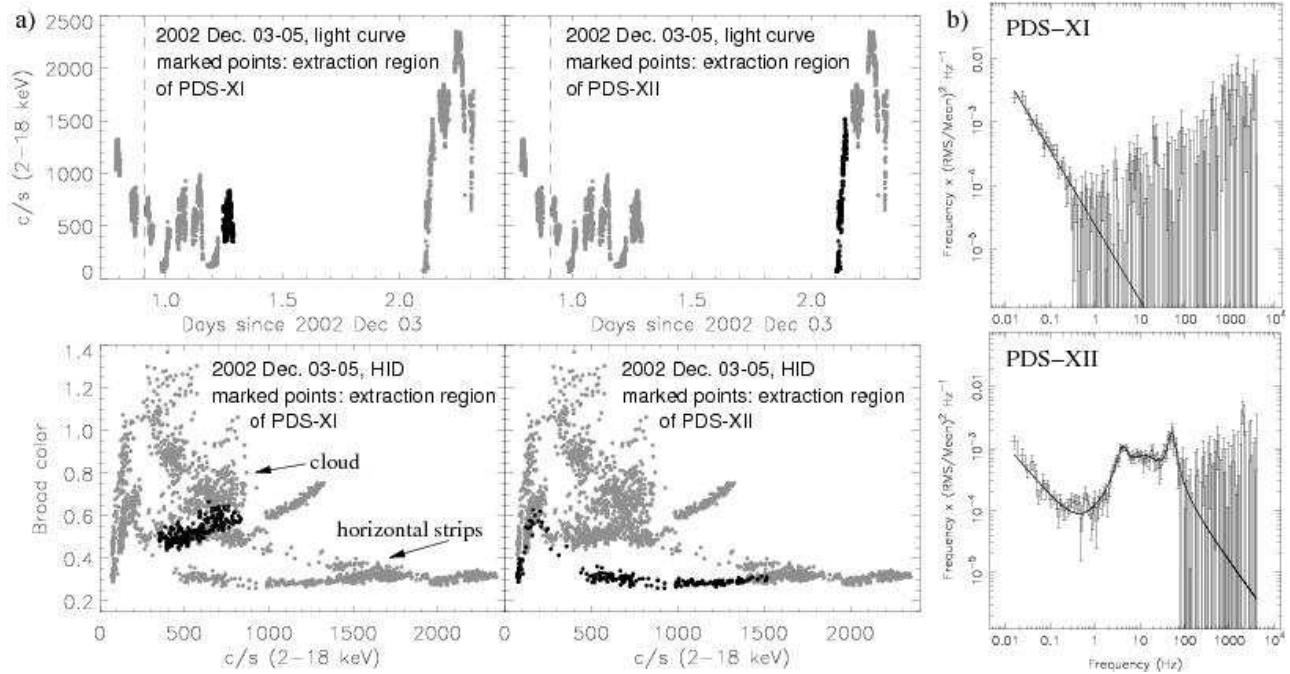}}
\end{tabular}
\caption{{\it a)} X-ray light curves and HIDs for intervals I and J. A vertical dashed line marks the periastron passage. ``Cloud'' and
``horizontal strips'' refer to two different regions mapped out by the source in the HID: the ``cloud'' corresponds to interval I and a few
points in interval J, the ``horizontal strips" are all drawn in interval J only. In the left-hand panels, the extraction region of PDS-XI
is marked with black points, both in the X-ray light curve and in the HID. In the right-hand panels we marked with black dots the
extraction region of PDS-XII. {\it b)} PDS-XI and PDS-XII, averaged on the marked data points in the panels on the left (a). Best fit parameters
are reported in Table \ref{tab:fit_pds}.}
\label{fig:hid_licu_pds_Dec2002}
\end{figure*}
We wonder whether the secular evolution of Cir X-1 X-ray flux might be responsible for these differences:
Shirey et al. (1998, 1999) observed the source in 1997, and its flux started decreasing from approximately years 1999-2000.
The X-ray flux in 2002 December was systematically lower than in 2000 October but we do not note any variation in the source spectral 
and timing behaviour. It follows that the different flux levels do not seem a satisfactory explanation for the differences with
Shirey et al. (1998, 1999). Furthermore, Cir X-1 shows properties typical of bright atoll sources after the periastron passage both in 2000
October and 2002 December, while Oosterbroek et al (1995)
identified atoll-like power spectra during a very low-flux period (their observations were performed with EXOSAT in 1984, 1985 and 1986 see
Figure 1 in Parkinson et al. 2003). This strengthens the idea that the variations of the average X-ray flux do not constitute a satisfactory
explanation for the peculiar behaviour of this source.

We observed three sequences of radio flares, two of them associated with phase 0.0 of the orbital cycle and one with
phase 0.5 of the orbital cycle. While the former can be explained invoking enhanced accretion close to the periastron passage,
the latter has a less clear interpretation. Two possible scenarios have been suggested in Tudose et al (2008): in the first one, the wind
accretion enhances close to the apastron passage, due to the decrease of the relative velocity between the compact object and the
stellar wind. The second possibility is that flares close to phase 0.5 are related to some mechanism taking place in the disc,
without involving accretion from stellar wind. These processes could increase the mass accretion rate, subsequently triggering radio flares.
\begin{figure*}
\begin{tabular}{c}
\resizebox{18cm}{!}{\includegraphics[angle=-90]{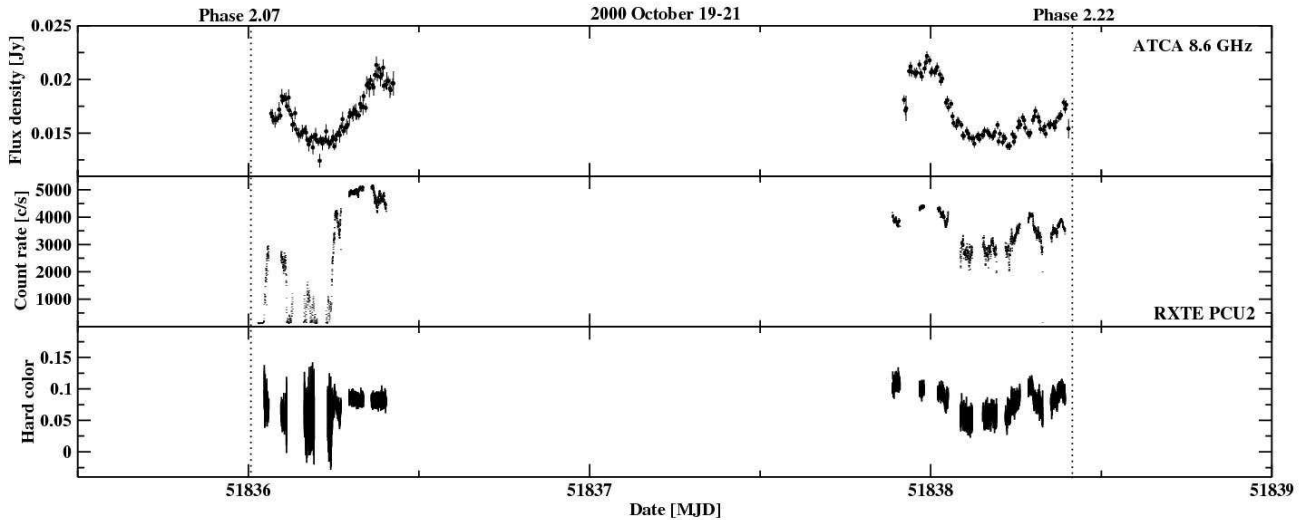}}
\end{tabular}
\caption{Simultaneous ATCA (8.6 GHz, top panel) and RXTE (PCU2 only, middle panel) light curves of Cir X-1 obtained on 2000
October 19-21 (intervals E and F). The HC curve is reported in the bottom panel. The bin size in the radio light curve is 5 min. The
vertical dotted lines mark two orbital phases, as indicated.}
\label{fig:radio_X_19_21_Oct}
\end{figure*}
From our data we can neither confirm nor rule out the two hypothesis.\\
On 2000 October 09-10, ATCA observed a radio flare close to phase 0.5 of the orbital cycle. We also observed an X-ray spectral
transition from the HB to the NB 2.5-3 hours after the end of the second radio flare (see \S\ref{par:0.5} and Figure
\ref{fig:licu_rad_hid_oct}). As we already pointed out, our data set does not allow to definitely say
whether the radio-flare sequence already ended at the end of the radio observation. Migliari et al. (2007) analysed simultaneous
X-ray/radio observation of the Z-source GX 17+2, reporting that a FB to NB transition precedes a resumption of the radio jet activity
by about 2 hours. They also showed that a radio jet forms after the FB to NB transition only when the NBO characteristic frequency
stabilizes in the power spectrum. This is particularly interesting: in an attempt to draw analogies between the properties
of low-frequency QPOs observed in neutron star Z-sources and black hole candidates (BHCs), Casella et al. (2005) associated
the FBO to the so-called type-A QPO,
the NBO to the type-B QPO and the HBO to the type-C QPO (Wijnands et al. 1999; Remillard et al. 2002). Casella et al. (2005) also
suggested that the HBO/type-C QPO and the NBO/type-B QPO could have respectively  similar production mechanisms. In BHCs the
type-B QPO is considered a signature of the spectral transition between hard-intermediate to soft-intermediate states (HIMS to SIMS, see
Homan \& Belloni 2005 for a state definition) as observed for example in GX 339-4 (Belloni et al. 2005) and XTE J1859+226
(Casella et al. 2004). These transitions might be in turn associated with the emission of relativistic radio-jets (Gallo et al. 2004;
Fender, Belloni \& Gallo 2004, Fender, Homan \& Belloni 2009). A type-B QPO has also been detected in GRS 1915+105 during SIMS to HIMS
transitions (Soleri et al. 2008a) and in H1743-322 in the HIMS to the low/hard state transition (Kalemci et al. 2006; the authors did
not classify the QPOs by type), when the radio-jet activity renews.
These results for BHCs together with the findings for the NS Z-source GX 17+2 suggest that there is a link between the type-B QPO/NBO
and the jet activity. In our data set, in 2000 October, the NBO appears $\sim 8$ hours after the beginning of the first radio
flare, assuming that the flares started at the beginning of the radio observation. Since the previous ATCA and RXTE observations
(interval B, HB) ended $\sim 2$ days before the beginning of observations of interval C, we can not be sure neither that the NBO did not
appear before the beginning of the radio flare nor that the radio flare really started at the beginning of interval C. From the analysis of
our data we can only observe that a spectral transition HB-NB occurs $\sim 8$ hours after the beginning of the first detected radio flare. We also
observe that a NBO appears, when the radio jet is possibly still on. Spectral variations in the X-ray band following radio flares have recently been
reported by Fender, Homan \& Belloni (2009) for the BHC GX 339-4. This is at variance with other sources, in which radio flares usually follow
X-ray spectral changes.\\
On 2002 December 05 we also observed a sequence of radio flares associated with the periastron passage. The source changes its position
in the X-ray HID after the flaring activity started. The shape of the PDS changes as well: PDS-XI only features VLFN, while in PDS-XII,
besides the VLFN, three Lorentzians are needed in order to fit the broad HFN, as already discussed for PDS-IX and PDS-X.
Inspecting Figure \ref{fig:licu_rad_dec} (top panel) we can see that the radio flux at the end of the observation simultaneous to
interval I is comparable to the radio flux of the observation simultaneous to interval J, when the flares are detected. This renders some
credibility to the possibility that in interval J we caught the beginning of the flare sequence. The HIDs and the light curves in
Figure \ref{fig:hid_licu_pds_Dec2002}a also suggest that the X-ray spectral transition from the cloud to the horizontal strips might have
taken place when the radio flare had already started, with a delay of $\sim 4.5$ hours.
This would be again similar to what has been reported for the BHC GX 339-4. As we noticed in \S\ref{par:Dec2002} the transition does not
happen at once: we can not associate the transition strips-cloud-strips to any specific re-flaring in the radio light curve.\\
It is interesting to note that in Cir X-1 we did not see any X-ray spectral transition preceding the launch of a radio jet (but in both cases
X-ray observations right before the onset of the radio flare are not available, so we might have missed something), as observed in GX
17+2 and as expected in the jet model presented in Migliari \& Fender (2006). We observed it twice in the other way around, with the radio
flares preceding the X-ray spectral changes, as also reported for the BHC and jet source GX 339-4 (Fender, Homan \& Belloni 2009).
It is natural to think that an X-ray spectral change reflects a change in the accretion conditions, and this will be reflected, with a time
delay, in the outflow, so in the jet. A possible explanation is that in Cir X-1 the accretion conditions are strongly ruled by the orbital
phase, since the orbit is probably particularly eccentric, and ``standard'' disc-jet connections do not apply as in other sources. Another
possible explanation is that changes in the accretion properties can not trigger outflows in a straightforward cause-effect relation, and that
could also explain the behaviour observed in GX 339-4.\\
ATCA also caught a radio flare at phase 0.0 on 2000 October 19-21. In this case we probably missed the beginning of the radio flare.
Also, RXTE could not detect any spectral transition, although the path drawn in the HID during interval A (note that the two binary orbits are not 
contiguous) suggests that a similar transition to the one detected on 2002 December 03-05 might have occurred. Differently from 2002 December
05, here the X-ray and the radio light curves seem to follow a similar trend. Typical power spectra
averaged on interval E are illustrated in PDS-VIII, -IX and -X. This similarity between the X-ray and the radio light curves is
interesting and completely different from what has been observed on 2002 December 03-05 and difficult to explain in the same framework.

\section{Conclusions}
We analysed simultaneous X-ray (RXTE) and radio (ATCA) observations of Cir X-1, performed in 2000 October and 2002 December. We found the
following:
\begin{enumerate}
 \item We identified typical Z-source features in the PDS as well as Z patterns drawn in an X-ray HID. At variance with Shirey et al. (1999),
 when close to the periastron passage, we could not find Z branches in the HID. PDS averaged after the periastron passage are reminiscent of the
 ones observed in some bright atoll sources that spend most of their time in the banana state. This is interesting, since in
 Cir X-1 atoll properties are detected  when the mass accretion rate $\dot{m}$ is particularly high (and not low, as expected), after phase 0.0 of
 the orbital cycle.
 \item some areas of the HID where Cir X-1 spends time before the periastron passage (the ``cloud'') can be classified neither in terms of Z-source
 nor in terms of atoll-source behaviour.
 \item On 2000 October 09-10, an X-ray spectral transition HB-to-NB is observed $\sim$8 hours after the beginning of the flaring sequence,
 $\sim2.5-3$ hours after the end of the second flare. We do not associate the transition and the NBO to
 any new re-flaring. Radio flares that precede X-ray spectral changes have also been seen in the BHC GX 339-4.
 \item On 2002 December 03-05, an X-ray spectral transition follows the beginning of a radio-flaring sequence with a delay of approximately
 4.5 hours. The transition occurs between two different areas of the HID (the cloud and the horizontal strips), characterized by a different
 shape of the light curve. The PDS also changes: before the transition a VLFN is the dominant component, after the transition the PDS shows
 a broad HFN reminiscent of what has been seen in several bright atoll sources.
 \item On 2000 October 19-21 we observe a flaring sequence in radio. Similar flares are also seen in the X-ray band. At variance with our
 observations on 2000 October 09-10 and 2002 December 03-05, where the radio flares precede any X-ray spectral change, here the two phenomena
 seem to follow a similar behaviour. This has been proposed in the toy model of Migliari
 \& Fender (2006) and observed in many BHCs (Fender, Belloni \& Gallo 2004), but not in the prototypical BHC and jet source GX 339-4
 (Fender, Homan \& Belloni 2009).
\end{enumerate}

\section*{Acknowledgments}
PS acknowledges support from a NWO VIDI awarded to R. Fender and a NWO Spinoza awarded to M. van der Klis. The Australia Telescope is funded by
the Commonwealth of Australia for operations as a national facility managed by CSIRO. This research has made use of data obtained through the
High Energy Astrophysics Science Archive Research Center Online Service, provided by the NASA Goddard space flight center.

\bsp

\label{lastpage}

\end{document}